\documentclass[a4paper, 12pt]{article}

\usepackage{amsfonts}
\usepackage{amssymb}
\usepackage{amsmath}

\usepackage[sort&compress,numbers]{natbib}
\usepackage{latexsym}
\usepackage{color}
\usepackage{braket}
\usepackage[colorlinks=true]{hyperref}
\usepackage{authblk}

\usepackage{graphicx,tikz}

\usepackage{amsthm}

\usepackage[titletoc]{appendix}

\addcontentsline{toc}{section}{Appendices}

\def\be{\begin{equation}}
\def\ee{\end{equation}}
\def\bea{\begin{eqnarray}}
\def\eea{\end{eqnarray}}
\def\bean{\begin{eqnarray*}}
\def\eean{\end{eqnarray*}}

\def\square{\hfill\hbox{\vrule\vbox{\hrule\phantom{N}\hrule}\vrule}\,}

\def\espaitemps{(M,g)}
\def\varietat{M}

\newtheorem{theorem}{Theorem}[section]

\newtheorem{defi}{Definition}

\newtheorem{remark}{Remark}[section]

\def\proof{\noindent{\em Proof.\/}\hspace{3mm}}

\newlength{\cellwidth}
\newcounter{mnotecount}[section]

\renewcommand{\themnotecount}{\thesection.\arabic{mnotecount}}
\newcommand{\mnote}[1]
{\protect{\stepcounter{mnotecount}}$^{\mbox{\footnotesize
$
\bullet$\themnotecount}}$ \marginpar{
\raggedright\tiny\em
$\!\!\!\!\!\!\,\bullet$\themnotecount: #1} }

\newcommand{\mnotex}[1]
{\protect{\stepcounter{mnotecount}}$^{\mbox{\footnotesize $\bullet$\themnotecount}}$
\marginpar{
\raggedright\tiny\em
$\!\!\!\!\!\!\,\bullet$\themnotecount: #1} }

\begin{document}

\title{Singularity theorems for warped products and the stability of spatial extra dimensions}


\author[1]{Nastassja Cipriani and Jos\'e M. M. Senovilla}
\affil[1]{Departamento de F\'isica Te\'orica e Historia de la Ciencia, Universidad del Pa\'is Vasco UPV/EHU, Apartado 644, 48080 Bilbao, Spain}

\maketitle

\vspace{-0.2em}

\begin{abstract}
New singularity theorems are derived for generic warped-product spacetimes of any dimension. The main purpose is to analyze the stability of (compact or large) extra dimensions against dynamical perturbations. To that end, the base of the warped product is assumed to be our visible 4-dimensional world, while the extra dimensions define the fibers, hence we consider {\em extra-dimensional evolution}. Explicit conditions on the warping function that lead to geodesic incompleteness are given. These conditions can be appropriately rewritten, given a warping function, as restrictions on the intrinsic geometry of the fibers ---i.e. the extra dimensional space. To find the results, the conditions for parallel transportation in warped products in terms of their projections onto the base and the fibers have been solved, a result of independent mathematical interest that have been placed on an Appendix.
\end{abstract}

\section{Introduction}
The fact that Einstein static universe is unstable was explicitly shown by Lema\^{\i}tre \cite{L1,L2} and Eddington \cite{E} many years ago. This is an old and well-known story which makes us wonder why Einstein did not realize this fact ---based on very basic physical arguments--- when confronted with Friedman's evolving solutions.\footnote{Curiously enough, when he eventually acknowledged the feasibility of these dynamical models is when he famously dismissed the cosmological constant as superfluous and probably unjustified \cite{Ein}.}

In a similar vein, and as a present to Stephen Hawking's 60th birthday, Penrose argued in 2002 that spatial compact extra-dimensions are likely to be unstable \cite{P}. He first recounted the ``wildly in excess'' number of degrees of freedom in higher-dimensional theories in comparison to what is perceived in ordinary physics.\footnote{He actually put forward a refined reasoning criticizing the ``usual string theorist's argument'' that the energy needed to excite the extra-space modes (particles) would be enormously large ---and thus experimentally inaccessible to us.}
He argued that the quantum numbers involved are large so that the proper way to approach the stability of compact extra dimensions is \underline{classically}. Then, the bulk of his argument to prove instability of extra compact dimensions was based on the fundamental singularity theorems proven by himself \cite{P1}, and later also with Hawking \cite{HP}, in the 1960's, see \cite{HE,S,SG}.

He ended up asserting \cite{P}:
\begin{quotation}
``(... a $(4+n)$-dimensional product spacetime) $M_4\times {\cal Y}$ is highly unstable against small perturbations. If ${\cal Y}$ is compact and of Planck-scale size, then {\em spacetime singularities} are to be expected within a tiny fraction of a second!''
\end{quotation}

To understand the reasoning behind this surprising claim, let us recall the classical Hawking-Penrose singularity theorem \cite{HP}.
\begin{theorem}[Hawking and Penrose 1970]\label{th:HP}
If the convergence, causality and generic conditions hold and if there is one of the following:
\begin{itemize}
\item a compact achronal set without edge,
\item a closed trapped surface,
\item a point with re-converging light cone,
\end{itemize}
then the space-time is causal geodesically incomplete.
\end{theorem}
Here, and for later use, we recall that the {\em convergence condition} is simply the requirement that
\be
R_{\mu\nu} v^\mu v^\nu \geq 0, \label{CC}
\ee
for arbitrary causal vectors $v^\mu$. If this is required only for null vectors then it is called the {\em null} convergence condition. The causality condition is the assumption that the spacetime is free from closed future-directed timelike curves. And the genericity condition implies that the geodesic deviation, ruled by
$$
R^\alpha{}_{\beta\mu\nu} u^\beta u^\mu ,
$$
is non-zero at least at a point of any causal curve with tangent vector $u^\mu$. As usual, $R^\alpha{}_{\beta\mu\nu}$ and $R_{\mu\nu}$ are the Riemann and Ricci tensors of the spacetime.

More importantly for our purposes is to understand the boundary condition requirement, that comes on three flavors. The first is simply a compact spacelike hypersurface without boundary (and without timelike related points). The third can be understood as the existence of a point whose future light cone \cite{P} `curls around and meets itself in all directions'. The second one will be further analyzed later in section \ref{sec:GStheorem}, when we discuss the concept of closed trapped submanifold in general.

\subsection{Penrose's argument}
But let us come back to Penrose's reasoning. To use the singularity theorems, he starts with a $(4+n)$-dimensional direct product $M_4\times {\cal Y}=\mathbb{R}\times\mathbb{R}^3\times {\cal Y}$ with metric as in  e.g.
\be
g = -dt^2 +dx^2+dy^2+dz^2 + g_{\cal Y},\label{flat+Y}
\ee
and \underline{perturbs} initial data given on a slice $\mathbb{R}^3\times {\cal Y}$ (say $t=0$) such that they do not `leak out' into the $\mathbb{R}^3$-part: they only disturb the ${\cal Y}$-geometry.
Letting aside the 3-dimensional typical large space represented by $\mathbb{R}^3$ one can consider a $(1+n)$-dimensional ``reduced spacetime'' $({\cal Z}, g_{red})$ whose metric $g_{red}$ is the evolution (for instance, a Ricci-flat solution) of perturbed initial data specified at ${\cal Y}$ ($t=0$). 
The full spacetime would be given by $\mathbb{R}^3\times {\cal Z}$ with direct product metric
$$
g_{pert}= g_{red}+dx^2+dy^2+dz^2 .
$$
But then, Theorem \ref{th:HP} applies to $({\cal Z}, g_{red})$ as it contains a compact slice and satisfies the convergence condition (because $R_{\mu\nu}=0$). 
He concluded that 
\begin{quotation}
``if we wish to have a chance of perturbing ${\cal Y}$ in a finite generic way so that we obtain a {\em non}-singular perturbation of the full $(4+n)$-spacetimes $M_4\times {\cal Y}$, then we must turn to consideration of disturbances that significantly spill over into the $M_4$ part of the spacetime''.
\end{quotation}
However, he claimed that such general disturbances are even more dangerous due to the large approaching Planck-scale curvatures that are likely to be present in ${\cal Y}$. He defended that there is good reason to believe that these general perturbations will also result in spacetime singularities, based again on Theorem \ref{th:HP}, but now using the third possibility: existence of a {\em point with reconverging light cone}.
 In the exact, unperturbed, models this of course fails as the models are non-singular, but adapting Penrose's writing
 \begin{quotation}
 ``...  {\it it just fails}. Only a `tiny' 2-dimensional subfamily of null geodesics generating the cone fail to wander into the ${\cal Y}$-part and back ---thus curling into the interior of the cone.

(...)

I believe that it is possible to show that with a generic but small perturbation (...) this saving property will be destroyed, so that the (...) singularity theorem will indeed apply, but a fully rigorous demonstration (...) is lacking at the moment. Details of this argument will be presented elsewhere in the event that it can be succinctly completed''.
\end{quotation}

These words were, {\it mutatis mutandis}, repeated in \cite{P2}. However, there has been no publication completing this argument since then. 

\subsection{Other arguments}
Almost simultaneously Carroll {\em et al} \cite{CGHW} argued that (large) extra dimensions must be dynamically governed by classical GR, and then showed that achieving static extra dimensions which are dynamically stable to small perturbations tends to be extremely difficult. They used a combination of the arguments based on singularity theorems with the existence of the stationary (or static) symmetry. Under the assumption of the mere null convergence condition the conclusion was that only cases with strictly positive Ricci curvature in all possible extra directions are feasible. In particular, flat extra-dimensional spaces are unstable, and even the addition of a cosmological constant does not seem to remedy this problem. Homogenous extra dimensions were assumed.
This type of instabilities was further analyzed in \cite{G}, where some arguments based on the weak cosmic censorship were put forward proving the existence of singularities ---which nevertheless might be hidden inside a black string.
 
 Since then, there have been several works analyzing this potential problem. For instance, in \cite{SW} how accelerated expansion imposes strong constraints on compact extra dimensions was discussed,
 implying either (i) that both the gravitational constant $G$ and the equation of state parameter $w$ depend on time, or (ii) violation of the null convergence condition (\ref{CC}) in an inhomogenous way across the extra dimensions together with a sinchronyzed variation with the observable matter and ``dark energy''. Their conclusions were nevertheless criticized in \cite{KP}, where one can find references to many other no-go and instability theorems \cite{KP}.

\subsection{This paper}
There are other physically motivated ways of probing the stability of spacetimes with extra dimensions, a classical key early result was given in \cite{W}, and rigorous PDE works on stability of product manifolds as solutions to the higher-dimensional Einstein equations under symmetry restrictions are given in \cite{Wyatt,BFK}, see also \cite{DFGHM}. In this paper, however, we want to concentrate on the arguments based on singularity, that is, incompleteness, theorems ---these were actually mentioned in \cite{CGHW} too. Penrose's reasoning is appealing, and whether or not it can be completed, or up to what extent used, is certainly intriguing. Our purpose is to, at least partly, provide an answer to this question. 

To that end, the difficulties to be confronted are of several kinds. First of all, as we have seen, the original argument by Penrose needed ---apart from underlying field equations to solve for the evolution--- some ad-hoc splittings. The reason behind is that the classical singularity theorem \cite{HP} is valid only for initial/boundary conditions placed on submanifolds of co-dimension 1, 2 or $D$ (here $D$ is the spacetime dimension). To address this problem we will use modern singularity theorems based on boundary conditions placed at submanifolds of {\em any possible} co-dimension derived some years ago in \cite{GS}, see also \cite{SG}. Section \ref{sec:GStheorem} is devoted to explaining these theorems and their underlying ideas, in particular the concept of trapped submanifold of arbitrary dimension, including closed trapped surfaces, will be studied.

A second major difficulty to be resolved is how to characterize generic but simple {\em geometrical} perturbations of a given stationary product spacetime $M_4\times {\cal Y}$. In this paper, we will consider the simplest geometrical perturbation one can think of: warping. Thus, the perturbed spacetime will be taken as a warped product with Lorentzian base $M_4$, fiber ${\cal Y}$ and warping function $f:M_4\rightarrow \mathbb{R}$. One could also consider a second possibility, taking ${\cal Y}$ as base and a warping function $f:{\cal Y} \rightarrow \mathbb{R}$
(this seems to be actually the case  called ``warped" whithin the string community). However, using a direct calculation, or general mathematical theorems on warped products, it is easy to prove that this kind of warping deformation is innocuous from our perspective. If the extra dimensional Riemannian part ${\cal Y}$ is compact
then the whole spacetime is geodesically incomplete if and only if the Lorentzian $M_4$ part is incomplete by itself \cite{RS}, see also \cite{CS}. In simpler words, compact extra dimensions do not change the (in)completeness of a given $M_4$ ---at least in this warped-product situation.

We will then concentrate on the mentioned possibility with $M_4$ as Lorentzian base, which contains in particular the cases with {\em dynamical} perturbations of the original direct-product manifold as then the function $f$ can depend on time. These are sometimes called {\em extra-dimension, or higher dimension, evolution} in the string community e.g. \cite{JJ}. In section \ref{sec:warped}, we will adapt the theorems in \cite{GS} to this situation and present singularity theorems applicable to general warped-product spacetimes. The essential assumptions of the theorems will be written in terms of properties of the Riemannian extra-dimensional space ${\cal Y}$ and of the Hessian of the warping function.

An analysis of the conditions of the theorems is then performed in section \ref{sec:analysis}. In particular, one can first of all derive some necessary condition on the sign of the Hessian of the warping function acting on particular timelike, or null, directions of the $M_4$-part. Assuming they hold, one can then rewrite the main condition in the theorems in the form of an inequality with quantities relative to the extra-dimensional space ${\cal Y}$ \underline{exclusively} on one side, and objects relative to the large 4-dimensional $M_4$ \underline{exclusively} on the other side ---formula (\ref{new1cond}) below. The former side has a controllable sign in many situations of physical interest. The importance of this particular form resides in that it provides a direct requirement to any thinkable extra space ${\cal Y}$ that one may wish to add to the visible 4-dimensional spacetime, discarding many {\em a priori} desirable possibilities. However, requirement (\ref{new1cond}) is too crude, and one can find an averaged condition of wider applicability and without the use of trapped submanifolds. This is the condition in Theorem \ref{new2}, or better its negation given in (\ref{notnew2}). Again the righthand side in (\ref{notnew2}) depends on the large 4-dimensional spacetime $M_4$ exclusively, and the lefthand side is an integral of a quantity essentially dependent of the extra-dimensional space ${\cal Y}$ ---apart from a positive factor $f^{-4}$. 
The outcome is that {\em dynamical} perturbations ---ruled by a warping function with timelike gradient--- of extra-dimensional spaces may be dangerous under some basic, physically motivated, assumptions ---in the sense that incomplete null geodesics may develop.

As a possible ``positive'' application of the theorems herein presented, and of conditions (\ref{new1cond}) and (\ref{notnew2}), one can argue that they may help in finding the stable possibilities. We stress that the results are valid both for compact and non-compact extra-dimensions, though the conditions must be placed on compact submanifolds of the extra-dimensional space, and thus the restrictions are stronger in the former case.

Some conclusions are gathered in section \ref{sec:conclusions}. We have added an Appendix of independent interest with results needed for the paper, where we solve the equations for parallel propagation in a warped-product semi-Riemannian manifold in terms of the projected equations to the base and the fibers. As far as we know, these results were not previously known.

\section{Singularity theorems based on submanifolds of arbitrary co-dimension}\label{sec:GStheorem}
In 1965, the first modern singularity theorem was presented in \cite{P1}. It was an important breakthrough in the field of Gravitation using for the first time the concept of geodesic incompleteness as indication of singular behaviour, and introducing the fundamental concept of closed trapped surface, see \cite{SG} for a recent review. We will use generalizations of this theorem below, so let us briefly remind how it goes.
\begin{theorem}[The 1965 Penrose singularity theorem]\label{th:P}
If the spacetime contains a non-compact Cauchy hypersurface and a closed trapped surface, and if the null convergence condition holds, then there exist incomplete null geodesics.
\end{theorem}

A Cauchy hypersurface is a spacelike hypersurface where good initial conditions can be placed, so that the whole future evolution can be determined from those given on it, basically because it is crossed once and only once by every causal curve \cite{HE,S,Wald}. 

In this theorem, the germinal and very fruitful notion of closed trapped surface was introduced. 
These are \underline{closed} spacelike surfaces (that is, compact without boundary co-dimension 2 spacelike submanifolds) such that their ``area'' ($(D-2)$-volume in general) tends to decrease locally along any possible {\em future} direction. There is of course a dual definition to the past.

As explained in the introduction, this concept was later used in the Hawking-Penrose Theorem \ref{th:HP}, together with the two other possibilities. The reasons why one needs to place the boundary condition of the singularity theorems on submanifolds of only co-dimension $1,2$ and $D$ were never explained, and seem at first sight unclear, especially taking into account that the property of being trapped can be trivially attached to submanifolds of arbitrary co-dimension $m$: it is enough to demand that its $(D-m)$-volume decreases locally along any possible future direction.

This question was addressed and clarified in \cite{GS}, where both Theorem \ref{th:HP} and Theorem \ref{th:P} were generalized in a neat way, allowing for trapped submanifolds of any co-dimension $m$. Let us briefly remind how these must be mathematically defined.

\subsection{Trapped submanifolds of arbitrary dimension}
To fix the notation, let $\espaitemps$ be a $D$-dimensional
Lorentzian manifold with metric tensor $g_{\mu\nu}$ of signature $(-,+,\dots,+)$. 
Consider an embedded spacelike submanifold $\zeta$ of any co-dimension $m$ and choose a basis $\{\vec{e}_A \}$ of vector fields tangent to $\zeta$ ($A,B,\dots =m+1,\dots , D$). Denote by $\gamma_{AB}$ the components of the (positive-definite) first fundamental form in the given basis: $\gamma_{AB}=g_{\mu\nu}|_\zeta e^{\mu}_Ae^{\nu}_B$.
Decomposing the derivatives of tangent vector fields $\{\vec{e}_A\}$ into its parts tangent and normal to $\zeta$ we have
$$
e^\rho_A\nabla_\rho e^\mu_B = \overline{\Gamma}^C_{AB}e^\mu_C-K^\mu_{AB},
$$
where $\overline{\Gamma}^C_{AB}$ provides the Levi-Civita connection of $\gamma_{AB}$ and $K^\mu_{AB}$ is called 
the shape tensor or {\em second fundamental form vector} of $\zeta$ in $\varietat$. It is symmetric in $AB$ and normal to $\zeta$ in its index $\mu$ by definition.

The mean curvature vector is defined simply as
$$H^\mu \equiv \gamma^{AB} K^\mu_{AB}. $$
Notice that $H^\mu$ is normal to $\zeta$. Therefore, it has $m$ independent components. These are usually called  \underline{expansions} of $\zeta$ relative to a chosen normal vector field $\vec n$ and are denoted and defined by
$$
\theta (\vec n) := n_{\mu}H^\mu . 
$$
If these expansions correspond to (future) {\em null} normals $\vec n$ (for the case with $m>1$), they are called (future) null expansions.

\begin{defi}[Trapped submanifold]
A spacelike submanifold $\zeta$ is said to be future trapped\footnote{In what follows, we will occasionally use the abbreviation f-trapped for future-trapped.}  if the mean curvature vector $\vec H$ is timelike and future-pointing everywhere on $\zeta$, and similarly for past trapped. 
\end{defi}
This is obviously equivalent to the statement that the expansions are negative 
$\theta (\vec n)<0$ for \underline{every} future pointing normal $\vec n$.

%
%

\subsection{The parallel propagated projector $P^{\mu\nu}$}
\begin{figure}[!ht]
	\includegraphics[height=10cm]{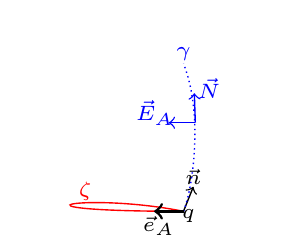}
\caption{Notation on a picture: Let the circle in red represent the closed spacelike submanifold $\zeta$. Then, pick up any future-pointing vector $\vec{n}$ orthogonal to $\zeta$ at a given point $q\in\zeta$ and launch the unique geodesic $\gamma$ tangent to $\vec{n}$ at $q$, represented here by the blue dotted line. The geodesic vector field tangent to $\gamma$ is then represented by $\vec{N}$. If $\{\vec{e}_A\}$ is a basis of vectors tangent to $\zeta$ at $q$,
 $\vec E_{A}$ denote the vector fields defined by parallel propagating $\vec{e}_{A}$ along $\gamma$. For further details, refer to the main text and \cite{GS}.\label{fig:notation}}
\end{figure}

To present the generalized singularity theorems some notation is needed. This is better understood on a picture, and thus explained in Figure \ref{fig:notation}. From now on, $n^{\mu}$ will denote a {\em future-pointing} normal vector to $\zeta$ at an arbitrary point $q\in \zeta$. Then $\gamma$ is the unique geodesic curve tangent to $n^\mu$ at $q\in \zeta$
with affine parameter $u$ along $\gamma$, we set $u=0$ at $q$, and denote by $N^{\mu}$ the vector field tangent to $\gamma$ (thus, $N^{\mu}|_{u=0}=n^{\mu}$). As above, $\{\vec{e}_A\}$ is a basis of vectors tangent to $\zeta$ at $q$
and $\vec E_{A}$ denote the vector fields defined by parallel propagating $\vec{e}_{A}$ along $\gamma$. Again, $\vec{E}_{A}|_{u=0}=\vec{e}_{A}$.

 By construction $g_{\mu\nu}E^\mu_{A}E^\nu_{B}$ are independent of $u$, so that 
 $$
 g_{\mu\nu}E^\mu_{A}E^\nu_{B}=g_{\mu\nu}e^\mu_{A}e^\nu_{B}=\gamma_{AB}
 $$
 all along $\gamma$. Then, along $\gamma$, we define the tensor field
 $$
 P^{\nu\sigma}:= \gamma^{AB}E^\nu_{A}E^\sigma_{B}, \hspace{1cm} P^{\nu\sigma}=P^{\sigma\nu}, \hspace{1cm} P^\nu{}_\nu = D-m .
 $$ 
 Observe that, at $u=0$, this is simply the projector to $\zeta$. Hence, $P^{\nu\sigma}$ is nothing but the parallel propagation of the $\zeta$-projector along $\gamma$.

%

\subsection{Generalized Penrose singularity theorems}
Armed with the presented notation, in \cite{GS} the following theorem was proven.
\begin{theorem}[Generalized Penrose singularity theorem]\label{genpen}
If $\espaitemps$ contains a non-compact Cauchy hypersurface and a closed f-trapped submanifold $\zeta$ of arbitrary co-dimension $m\geq 2$, and if
\be
\fbox{$R_{\mu\nu\rho\sigma}N^\mu N^\rho P^{\nu\sigma}\geq 0$} \label{cond}
\ee
holds along every future-directed null geodesic emanating orthogonally from $\zeta$, then $\espaitemps$ is future null geodesically incomplete.
\end{theorem}
In \cite{GS} the generalization of Theorem \ref{th:HP} was also achieved allowing for a closed set of arbitrary co-dimension by simply adding the condition (\ref{cond}). It must be remarked, in this sense, that
\begin{enumerate}
\item For spacelike hypersurfaces, co-dimension $m=1$, there is a unique timelike orthogonal direction $n^{\mu}$ initially. Then, letting $N^\mu$ denote its geodesic extension as explained in figure \ref{fig:notation}, 
$P^{\mu\nu}=g^{\mu\nu}-(N_{\rho}N^\rho)^{-1}N^{\mu}N^{\nu}$ and (\ref{cond}) reduces simply to 
$$R_{\mu\nu}N^\mu N^\nu \geq 0,$$
that is, the {\em timelike convergence condition} along $\gamma$. Thus, nothing was added to Theorem \ref{th:HP} in this case.
\item For spacelike `surfaces', co-dimension $m=2$, there are two independent null normals on $\zeta$, say $n^{\mu}$ and $\ell^{\mu}$. Define $L^{\mu}$ parallelly propagating $\ell^\mu$ along $\gamma$. Then, it is easily seen that $P^{\mu\nu}=g^{\mu\nu}-(N_{\rho}L^\rho)^{-1}(N^{\mu}L^{\nu}+N^{\nu}L^{\mu})$ and again (\ref{cond}) reduces to
$$R_{\mu\nu}N^\mu N^\nu \geq 0,$$
that is to say, the {\em null convergence condition} along $\gamma$. Again, nothing has been added to Theorem \ref{th:HP} in this case. And for Theorem \ref{genpen} observe that (\ref{cond}) is simply equivalent to the null convergence condition in this case. 
\item For points, co-dimension $m=D$, the situation is a little more involved. However, one can use a reformulation in terms of Jacobi tensors \cite[Prop. 12.46]{BEE} to put (\ref{cond}) in relation with a `genericity' condition $R_{\mu\nu\rho\sigma}N^\mu N^\rho\neq 0$.
\end{enumerate}
These three cases cover the original Hawking-Penrose Theorem \ref{th:HP}.
The physical and mathematical interpretation of condition (\ref{cond}) for co-dimensions other than $1,2$ or $D$ is given in terms of tidal forces, or equivalently in terms of sectional curvatures, and can be consulted in \cite{GS}.

The curvature condition (\ref{cond}) can in fact be substantially weakened, and it is sufficient that it holds on the average along $\gamma$ in a certain sense. This is remarkable, because it permits to prove singularity theorems without any trapped submanifold! Specifically, the conclusion of Theorem \ref{genpen} remains valid if the curvature condition (\ref{cond}) and the trapping condition $\zeta$ assumed there are jointly replaced by
$$
\int_0^{a} R_{\mu\nu\rho\sigma}N^\mu N^\rho P^{\nu\sigma} du >  \theta(\vec n)
$$
along each future inextendible null geodesic $\gamma : [0,a) \to \varietat$ emanating orthogonally from $\zeta$ with initial tangent $n^\mu$. This was explicitly proven in \cite{GS}. Here, we present a different but equivalent version, so that one does not need to assume anything along inextendible but incomplete null geodesics.
\begin{theorem}\label{genpen2}
If $\espaitemps$ contains a non-compact Cauchy hypersurface and is null geodesically complete, then for \underline{every} closed spacelike submanifold $\zeta$ of co-dimension $m>1$ there exists at least one null geodesic $\gamma$ with initial tangent $n^\mu$ orthogonal to $\zeta$ along which
\be
\fbox{$\displaystyle{\int_0^{\infty} R_{\mu\nu\rho\sigma}N^\mu N^\rho P^{\nu\sigma} du \leq  \theta(\vec n)}$} \, .\label{cond2}
\ee
\end{theorem}

Observe that there is no restriction on the sign of $\theta(\vec n)$, and thus this theorem can be applied to minimal submanifolds, or even to other cases with some positive initial expansions.

\section{Singularity theorems for warped-product spacetimes}\label{sec:warped}
Consider a direct product $(4+n)$-dimensional spacetime $M=M_4\times {\cal Y}$ with direct product metric
$$
g_{\mu\nu}dx^\mu dx^\nu= \hat{g}_{ab}(x^c)dx^a dx^b + \bar{g}_{ij}(x^k)dx^i dx^j
$$
where $x^\mu =(x^a,x^i)$ are local coordinates, those with indices 
$a,b,\dots ,h$ relative to the 4-dimensional $M_4$, and those with indices $i,j,k,l$ relative to the $n$-dimensional ${\cal Y}$. Note that the total dimension is $D:= 4+n$. Assume, of course, that the metric $\hat g$ on $M_4$ is Lorentzian and the metric $\bar g$ on ${\cal Y}$ is positive definite.

\subsection{Perturbation: warped products}
To make sense of the instability arguments presented in the introduction, we want to perturb these spacetimes geometrically. The simplest way to do this is by breaking the direct product structure and letting one of the two pieces influence the other via a warping function $f$. Now, there are two inequivalent ways of doing this. One could let the extra-dimensional part influence the large visible spacetime $M_4$ by placing a factor $f^2(x^i)$ in front of the Lorentzian metric $\hat{g}_{ab}(x^c)dx^a dx^b$. 
However, it is easily seen that this cannot lead to any incompleteness of geodesics. This follows for instance from a known mathematical theorem stating that if the Riemannian base of a warped product is complete ---in particular, this is always the case for a compact base---  and the warping function is bounded away from zero by a positive constant
then the entire manifold is geodesically complete if and only if the fiber so is \cite{RS,CS}\footnote{Notice that
the condition $f\geq \epsilon>0$ for a complete but non-compact base $({\cal Y},\bar g)$ is quite acceptable, even logical, in the given setting as we are perturbing the product case with $f=1$; furthermore, this assumption can be substantially relaxed, see Remark 3.11 in \cite{RS}.}. 
Hence, the influence of the extra dimensions via a warping function can never render the 4-dimensional visible spacetime incomplete: either $(M_4,\hat g)$ is incomplete by itself or not.

The other possibility is more interesting, as it includes {\em dynamical} perturbations of the extra dimensions. Thus, to fix ideas, let the  {\em warped product} $M_4\times_f {\cal Y}$ spacetime have Lorentzian base $M_4$, fiber ${\cal Y}$ and warping function $f: M_4 \rightarrow \mathbb{R}$, thus perturbing the original spacetime in the following manner:
\be
g_{\mu\nu}dx^\mu dx^\nu= \hat{g}_{ab}(x^c)dx^a dx^b + f^2(x^c) \bar{g}_{ij}(x^k)dx^i dx^j .\label{warped}
\ee
This type of metrics are able to describe ``higher-dimension evolutionary'' cases and have been considered in the literature to analyze the possibility of viable cosmological models with extra dimensions \cite{F,FR}, and also in connection with stability issues e.g. \cite{Maeda,G}, especially for functions $f$ depending on time, see also \cite{JJ}.

The components of the curvature tensor are readily computed (for instance from (\ref{gamma1}-\ref{gamma2}) in the Appendix) leading to 
\bea 
R^a{}_{ijk} &=&0, \hspace{1cm} R^i{}_{abc} =0, \hspace{1cm} R^i{}_{jab} =0, \label{Rie1}\\
R^a{}_{ibj} &=& -f \hat\nabla_b \hat\nabla^a f \, \bar{g}_{ij} , \\
R^i{}_{jkl} &=& \overline{R}^i{}_{jkl} -\hat\nabla^af \hat\nabla_a f \left(\delta^i_k \bar{g}_{jl}-\delta^i_l \bar{g}_{jk} \right), \\
R^a{}_{bcd} &=& \hat{R}^a{}_{bcd} \label{Rie4}
\eea
where $\overline{R}^i{}_{jkl} $ is the Riemann tensor of $({\cal Y},\bar g)$, while $\hat\nabla$ and $\hat{R}^a{}_{bcd}$ are the connection and curvature tensor, respectively, of $(M_4,\hat g)$. From here it is easy to get the Ricci tensor
\bea
R_{ab} &=& \hat{R}_{ab} -n \frac{1}{f} \hat\nabla_a \hat\nabla_b f,\label{ricci1}\\
R_{ai} &=& 0, \label{ricci2}\\
R_{ij} &=& \bar{R}_{ij} -\bar{g}_{ij} \left(f\hat\nabla^b\hat\nabla_b f +(n-1) \hat\nabla_b f \hat\nabla^b f \right)\label{ricci3}
\eea
where $\hat{R}_{ab}$ and $\overline{R}_{ij}$ are the respective Ricci tensors on the hat and bar parts.

\subsection{Warped products: null geodesics}\label{nullgeod}
The way geodesics behave in warped products, in relation with the projected curves, is well known, see for instance \cite{O}. We recall here the results needed to apply Theorem \ref{genpen} and Theorem \ref{genpen2} to the metric (\ref{warped}).
Let $\gamma: x^\mu =x^\mu (u)$ be an affinely parametrized \underline{null} geodesic with tangent vector $dx^\mu/du:= N^\mu =(\hat{N}^a,\bar{N}^i)$, so that $\hat{N}^a:=dx^a/du$ and $\bar{N}^i := dx^i/du$ are the vectors tangent to the respective projections of $\gamma$ into $M_4$ and ${\cal Y}$. Then
\be
\bar{N}^j\overline\nabla_j \bar{N}^i = -2  \hat{N}^a\partial_a (\ln f) \, \bar{N}^i=-2\frac{d\ln f|_\gamma}{du} \bar{N}^i, \label{nonaffine}
\ee
which actually states that the curve projected to ${\cal Y}$ is itself a geodesic though non-affinely parametrized. From this one deduces that 
\be
\bar{g}_{ij} \bar{N}^i\bar{N}^j = \frac{C}{f^4}, \hspace{1cm} C =\mbox{const.}, \label{C}
\ee
where $C\geq 0$ is a non-negative constant. In particular, if the ${\cal Y}$-initial velocity vanishes $\bar{N}^i (0)=n^i=0$, then $\bar{N}^i(u)=0$ for all $u$. Hence, clearly $C=0$ means that the null geodesic $\gamma$ lives exclusively in the Lorentzian part $(M_4,\hat g)$ of the warped product.

For the other projected curve, one has on using that $\gamma$ is null 
\be
\hat{N}^b\hat\nabla_b \hat{N}^a = -(\hat{g}_{bc} \hat{N}^b\hat{N}^c) \, \hat\nabla^a (\ln f). \label{subgeod}
\ee
This tells us that the acceleration of the $M_4$-projected curve is always parallel to the gradient of $f$. Such curves are called {\em subgeodesics with respect to $\nabla f$}.
Observe that this equation involves only quantities of the $(M_4,\hat g)$ part, and thus its solutions are well-defined by giving initial conditions on that part: it is a good, well defined, transport equation. 

By using $g_{\mu\nu}N^\mu N^\nu =0$ and (\ref{C}) it is immediate to show that
\be
\hat{g}_{ab} \hat{N}^a \hat{N}^b = -\frac{C}{f^2}, \label{C1}
\ee
which permits to rewrite (\ref{subgeod}) simply as
\be
\hat{N}^b\hat\nabla_b \hat{N}^a = \frac{C}{f^3} \hat\nabla^a f =-\frac{C}{2} \hat\nabla^a\left(\frac{1}{f^2} \right). \label{subgeod2}
\ee
This equation can be analyzed as that of a particle moving on a potential $\displaystyle{V(\hat x) =\frac{C}{2f^2(\hat x)}}$ and then (\ref{C1}) is a first integral stating that the particle has ``vanishing total energy''. 

To get the information contained in these formulas one can proceed as follows: given that (\ref{subgeod}) ---or equivalently (\ref{subgeod2})--- is a good transport equation one starts by solving it with given initial conditions. The solution provides the part of the geodesic projected to $M_4$, that is $\hat\gamma : \hat x^b (u)$. As $\hat x^b (u)$ are then known explicitly, the function $f$ is also known along $\gamma$, given by $f(\hat x^b(u))$. The next step is simply to solve the geodesic equation (\ref{nonaffine}). A convenient way of expressing the solution is
\be
\bar N^i = f^{-2}|_\gamma {\cal N}^i , \hspace{1cm} {\cal N}^i (0) = f^2(0) n^i, \label{calN}
\ee
where ${\cal N}^i$ is the affinely parametrized geodesic vector field in $({\cal Y},\bar g)$ with initial condition as stated. Note that 
$$
C = \bar{g}_{ij} {\cal N}^i(0) {\cal N}^j(0) \, \, (=f^4(0) \bar{g}_{ij} n^i n^j )
$$
so that, whenever $C\neq 0$, it can be set equal to $º1$ by choosing the initial $n^i$ appropriately. Furthermore, observe that the relation of the affine parameter $\bar u$ of the projected geodesic $\bar\gamma$ with the affine parameter $u$ of the spacetime geodesic $\gamma$ is given by
\be
du = f^2|_\gamma d\bar u .\label{baru}
\ee

\subsection{Parallel transport along null geodesics orthogonal to $\zeta$}
\label{ParallelTransport}
 As $(M_4,\hat{g})$ is Lorentzian, we are going to choose our test submanifolds ``living'' on the extra-dimensional part ${\cal Y}$, that is to say, they will be the lifts to the spacetime of a given submanifold $\zeta\subset {\cal Y}$ which is compact and with total co-dimension $m$. In other words, our submanifolds will have constant values of the coordinates $x^b$. 
In order to fix ideas, some notation must be used. We will denote by ${\cal Y}_p$ any ``copy'' of ${\cal Y}$ in the manifold: more precisely, let $p\in M_4$ be any point of the Lorentzian 4-dimensional manifold, say given by coordinates $x^a(p) =\hat x^a_p$. Then, the submanifold ${\cal Y}_p\subset M$ is defined by 
$$
{\cal Y}_p := \{ x^a =\hat x^a_p\},
$$
which as a topological manifold is simply ${\cal Y}_p =\{p\}\times {\cal Y}$. Now, any given submanifold $\zeta\subset {\cal Y}$ automatically defines a submanifold $\zeta_p \subset {\cal Y}_p$ for any $p\in M_4$ in the straightforward way.
 
Let $\{\vec{e}_A\}$ denote an ON basis of vector fields tangent to $\zeta_p$: $e^\mu_A =(0,\bar{e}^i_A)$.
Using the general results in the Appendix, one can easily obtain that along any null geodesic orthogonal to $\zeta_p$ we have
$$E_A^\mu =(0,\bar{E}^i_{A\parallel}/f)$$
where $\bar{E}^i_{A\parallel}$ are the parallel transports of $f|_p \bar{e}^i_A$ along the curve projected to ${\cal Y}_p$ given by $\overline{\gamma}: x^i(u)$. Explicitly
$$\bar{N}^j\overline\nabla_j \bar{E}^i_{A\parallel} =0, \hspace{1cm} \bar{E}^i_{A\parallel} (0) =f|_p \bar{e}^i_A . $$

From this expression, or alternatively recalling that parallel transportation respects scalar products, one also deduces
 \bean
 g_{\mu\nu}N^\mu E^\nu_A =0 \hspace{4mm} \Longrightarrow \hspace{5mm} \bar{g}_{ij}\bar{N}^i \bar{E}^j_{A\parallel}=0,\\
 g_{\mu\nu}E_B^\mu E^\nu_A =\delta_{BA} \hspace{4mm} \Longrightarrow \hspace{5mm} \bar{g}_{ij}\bar{E}_{A\parallel}^i \bar{E}_{B\parallel}^j= \delta_{AB}.
 \eean
 It follows that the tensor $P^{\mu\nu}=\gamma^{AB} E^\mu_A E^\nu_B$ becomes simply along $\gamma$
$$
P^{ab} =0, \hspace{4mm} P^{ia} =0, \hspace{4mm} P^{ij} =\frac{1}{f^2} \delta^{AB}  \bar{E}^i_{A\parallel}  \bar{E}^j_{B\parallel}.
$$
Introducing this into the lefthand side of condition (\ref{cond}) of Theorem \ref{genpen} and using (\ref{Rie1}-\ref{Rie4}) for the Riemann tensor, a little calculation leads to \cite{Cipriani}
$$R_{\mu\nu\rho\sigma}N^\mu N^\rho P^{\nu\sigma} =\delta^{AB}\bar{R}_{ijkl}\bar{N}^i\bar{N}^k \bar{E}^j_{A\parallel} \bar{E}^l_{B\parallel}
 -(D-m) \frac{1}{f} \hat N^a\hat\nabla_a (\hat N^b\hat\nabla_b f) |_\gamma .$$
The last summand here is proportional to the second derivative of the warping function $f|_\gamma$ along $\gamma$ with respect to the affine parameter $u$: 
$$\hat N^a\hat\nabla_a (\hat N^b\hat\nabla_b f) |_\gamma =
\displaystyle{\frac{d^2 f|_\gamma }{du^2}}.
$$
The first summand can be alternatively expressed in terms of the affine geodesic vector field ${\cal N}^i$ defined in (\ref{calN}), rendering the expression in the form
\be
R_{\mu\nu\rho\sigma}N^\mu N^\rho P^{\nu\sigma} =\left.\frac{1}{f^4} \delta^{AB}\bar{R}_{ijkl}{\cal N}^i{\cal N}^k \bar{E}^j_{A\parallel} \bar{E}^l_{B\parallel}-(D-m) \frac{1}{f} \frac{d^2 f}{du^2}\right|_\gamma . \label{newcond}
\ee

\subsection{The singularity theorems adapted to warped products}
Expression (\ref{newcond}) is all that we need to produce new singularity theorems as corollaries of Theorem \ref{genpen} and Theorem \ref{genpen2} adapted to the warped product situation (\ref{warped}).\footnote{Singularity theorems applicable to warped-product spacetimes with Lorentzian base have been previously found in e.g. \cite{C,WW}. Our theorems are of a different nature, as they assume non-compact Cauchy hypersrfaces, and more importantly, they use as boundary condition submanifolds of arbitrary co-dimension.} With the above notation at hand and starting with Theorem \ref{genpen}, use of (\ref{newcond}) allows one to derive the following theorem.
\begin{theorem}\label{new1}
Let $\espaitemps$ be a warped product $M=M_4\times_f {\cal Y}$ with metric (\ref{warped}) that possesses a non-compact Cauchy hypersurface and a closed f-trapped submanifold $\zeta_p \subset {\cal Y}_p$ ($p\in M_4$) of co-dimension $m$. For any normal vector $n^i \in T_q{\cal Y}_p$ orthogonal to $\zeta_p$ (including the zero vector) let $\bar\gamma (\bar u)$ be the affinely parametrized geodesic in $({\cal Y}_p,\bar g)$ tangent to $n^i$ at $q\in \zeta_p$ with tangent vector field ${\cal N}^i$ and affine parameter $\bar u$. Let also $\bar{E}^i_{A\parallel}$ denote the parallel transports in $({\cal Y}_p,\bar g)$ along $\bar\gamma(\bar u)$ of the elements of an orthonormal basis $\{e^i_A\}$ of vectors tangent to $\zeta_p$ at $q\in \zeta_p$. 

Let $\hat\gamma(u): x^a(u)$ with $\hat N^a =dx^a(u)/du$ denote any solution of the equation (\ref{subgeod2}) with initial condition $\hat N^a|_p= n^a$, where $n^a$ is such that $\hat{g}_{ab}|_p n^a n^b=-f^2|_p \bar{g}_{ij} n^i n^j$ and future pointing. If the following inequality (of functions of $\bar u$)
\be
\delta^{AB}\bar{R}_{ijkl}{\cal N}^i{\cal N}^k \bar{E}^j_{A\parallel} \bar{E}^l_{B\parallel}>
\left. -(D-m) f^2 \frac{d^2 f^{-1}}{d\bar u^2} \right|_{\hat\gamma} \label{new1cond}
\ee
holds for each $n^i$ at all $q\in \zeta_p$, and for all possible choices of $n^a$ accordingly, then $\espaitemps$ is future null geodesically incomplete.
\end{theorem}
\proof We only have to check that this is equivalent to Theorem \ref{genpen}. Obviously, the assumptions are the same, so that only the equivalence of (\ref{cond}) with (\ref{new1cond}) must be justified. And this follows from the analysis in subsection \ref{ParallelTransport}, specifically from expression (\ref{newcond}) (times $f^4$), which gives immediately the lefthand side of (\ref{new1cond})  together with the following straightforward calculation for the remaining term in (\ref{newcond}), containing derivatives of $f$ along $\hat\gamma$, on using (\ref{baru})
$$
\frac{d^2 f}{du^2}=\frac{1}{f^4}\frac{d^2f}{d\bar u^2}-\frac{2}{f^5} \left(\frac{df}{d\bar u} \right)^2=-\frac{1}{f^2} \frac{d^2 f^{-1}}{d\bar u^2}.
$$
\square

\begin{remark}
We have chosen to express the main condition in the above theorem in the form (\ref{new1cond}) because the lefthand side involves a quantity intrinsic to the extra dimensional space $({\cal Y},\bar g)$, while the righthand side depends exclusively on the behaviour of the warping function along the timelike subgeodesics (or null geodesics if $n^i=0=C$)  defined by (\ref{subgeod2}). Thus, the righthand side depends exclusively on the Lorentzian manifold $(M_4,\hat g)$ ---and $f$. 
\end{remark}

\begin{remark}\label{C=0}
The above theorem requires that the condition holds also for the null geodesics with $n^i=0$, which is equivalent to $C=0$. As explained previously, these are the null geodesics with no component in (or trivial projection to) the extra-dimensional space ${\cal Y}$, $N^i =0$. 
In this situation there is no need, and actually it makes no sense, to define the parameter $\bar u$, so that the condition should be written in terms of the null geodesic affine parameter $u$ and reads simply from (\ref{newcond})
\be
\frac{d^2f}{du^2} <0 \hspace{1cm} \mbox{if} \hspace{3mm} C=0.\label{new1cond'}
\ee
This must hold for all such null geodesics orthogonal to $\zeta_p$. Given that such null geodesics always exist (there is actually a 3-parameter family of them), this must be considered as a necessary condition for Theorem \ref{new1} to hold.
\end{remark}

An analysis of the above conditions (\ref{new1cond}) and (\ref{new1cond'}) is left for the next section. Let us turn to the stronger Theorem \ref{genpen2}. The condition in the theorem involves an integral along the spacetime null geodesic and this is compared to the initial expansion. Therefore, we need to compute this initial expansion for any closed $\zeta_p\in {\cal Y}_p$. A simple direct  computation gives, for the initial expansion along $\vec n$:
\be
\theta(\vec n) = \bar{\theta}_{\bar{n}} +(D-m) \left.\frac{1}{f} n^a\partial_a f\right|_p, \label{thetan}
\ee
where $\bar{\theta}_{\bar{n}}$ is the ``expansion of $\zeta_p$ as submanifold of $({\cal Y}_p,\bar{g})$'': compute the mean curvature vector $\bar H^i$ of $\zeta_p$ as a submanifold of ${\cal Y}_p$, and then $\bar\theta_{\bar n} := \bar g(\bar H,\bar n)=\bar g_{ij}\bar H^in^j$. The above displayed expression gives the righthand side of the condition (\ref{cond2}) in Theorem \ref{genpen2}. To find the lefthand side, it is enough to use (\ref{newcond}). We thus have
\begin{theorem}\label{new2}
Let $\espaitemps$ be a warped product $M=M_4\times_f {\cal Y}$ with metric (\ref{warped}) that possesses a non-compact Cauchy hypersurface and is null geodesically complete. Let $\zeta_p \subset {\cal Y}_p$ ($p\in M_4$) denote a closed submanifold of co-dimension $m$, and for any normal vector $n^i \in T_q{\cal Y}_p$ orthogonal to $\zeta_p$ (including the zero vector) let $\bar\gamma (\bar u)$ be the affinely parametrized geodesic in $({\cal Y}_p,\bar g)$ tangent to $n^i$ at $q\in \zeta_p$ with tangent vector field ${\cal N}^i$ and affine parameter $\bar u$. Let also $\bar{E}^i_{A\parallel}$ be the parallel transports along $\bar\gamma(\bar u)$ in $({\cal Y}_p, \bar g)$ of the elements of an orthonormal basis $\{e^i_A\}$ of vectors tangent to $\zeta_p$ at $q\in \zeta_p$. 

Let $\hat\gamma(u): \hat x^a(u)$ with $\hat N^a =d\hat x^a(u)/du$ denote any solution of the equation (\ref{subgeod2}) with initial condition $\hat N^a|_p= n^a$, where $n^a$ is such that $\hat{g}_{ab}|_p n^a n^b=-f^2|_p \bar{g}_{ij} n^i n^j$ and future pointing. To relate both curves $\bar\gamma(\bar u)$ and $\hat\gamma(u)$ ---so that they provide a null geodesic in the spacetime--- set, as in (\ref{baru}), $du=f^2|_{\hat\gamma} d\bar u$. 

Then, for \underline{every} such closed $\zeta_p \subset {\cal Y}_p$ there exists at least one choice of $n^i$ and $n^a$ at some $q\in \zeta_p$ such that the following inequality holds
\bean
\int^\infty_0 \frac{1}{f^4}\delta^{AB}\overline{R}_{ijkl}{\cal N}^i {\cal N}^k \bar{E}^j_{A\parallel} \bar{E}^l_{B\parallel}du -\bar{\theta}_{\bar{n}}
 \leq   (D-m) \left(\int_0^\infty \left.\frac{1}{f} \frac{d^2 f}{du^2}\right|_{\hat\gamma} du + \frac{1}{f(p)} n^a\partial_a f|_p \right)\, .
\eean
\end{theorem}
\proof This is just a corollary of Theorem \ref{genpen2} by using (\ref{newcond}), (\ref{thetan}) and rearranging, together with the results in subsections \ref{nullgeod} and \ref{ParallelTransport}.\square

Again the condition is written in a form that is considered optimal. The righthand side involves only quantities of the large 4-dimensional spacetime $(M_4,\hat g)$ and the warping function $f:M_4\rightarrow \mathbb{R}$. The lefthand side is not completely intrinsic to the extra-dimensional space $({\cal Y},\bar g)$ due to the factor $1/f^4$. Still, this factor is strictly positive in all cases, which allows one to control its influence up to some degree, and concentrate on the analysis of the main integrand term $\delta^{AB}\overline{R}_{ijkl}{\cal N}^i {\cal N}^k \bar{E}^j_{A\parallel} \bar{E}^l_{B\parallel}$, which is a quantity intrinsic to $({\cal Y},\bar g)$. Observe that there are several ways to write the integral on the lefthand side of the theorem's condition, viz.
\bea
\int^\infty_0 \frac{1}{f^4}\delta^{AB}\overline{R}_{ijkl}{\cal N}^i {\cal N}^k \bar{E}^j_{A\parallel} \bar{E}^l_{B\parallel}du=
\int^\infty_0 \delta^{AB}\overline{R}_{ijkl}N^i N^k \bar{E}^j_{A\parallel} \bar{E}^l_{B\parallel}du \label{form1}\\
= \int^{\bar u_{\infty}}_0 \frac{1}{f^2}\delta^{AB}\overline{R}_{ijkl}{\cal N}^i {\cal N}^k \bar{E}^j_{A\parallel} \bar{E}^l_{B\parallel}d\bar u \label{form2}\\
=\int^{\bar u_{\infty}}_0 \delta^{AB}\overline{R}_{ijkl}{\cal N}^i {\cal N}^k \bar{E}^j_{A} \bar{E}^l_{B}d\bar u, \label{form3}
\eea
where $\bar u_\infty$ is the value of $\bar u$ as $u\rightarrow \infty$. Still, none of them is intrinsic to $({\cal Y},\bar g)$. In (\ref{form1}) the warping function has disappeared {\em explicitly} but it remains there implicitly as it is necessary to construct the vector field $N^i$ along $\bar\gamma$ by means of (\ref{nonaffine}), or directly from (\ref{calN}). A similar comment applies to (\ref{form3}), as $f$ is needed to compute the vector fields $\bar E^i_A$, and we also need to know $f$ to compute $\bar u_\infty$ via (\ref{baru}). This also happens in (\ref{form2}) where, in addition, $f$ appears explicitly. 

To end this subsection we wish to remark that all of the above has been done to the future, but one can also derive the correspoding dual versions to the past: it is enough to choose $n^a$ past-directed ---and, in the case of Theorem \ref{new1}, that $\zeta_p$ is {\em past}-trapped.

\subsection{Null geodesic incompleteness manifestation}
The theorems of the previous subsection prove null geodesic incompleteness under some conditions to be analyzed later in the next section. Before that, we want to understand how this incompleteness can arise for a metric of type (\ref{warped}). The answer is that such incompleteness can only happen if {\em either the warping function $f$ or its inverse $1/f$ approach zero somewhere on the Lorentzian base $(M_4,\hat g)$}. 

This can be deduced from a combination of some interesting results in \cite{CS,RS}. In particular, adapted to our situation, we have
\begin{theorem}[\cite{CS,RS}] \label{sanchez}
If $0<\epsilon \leq f \leq A$ for constants $\epsilon$ and $A$, then the warped product (\ref{warped}) with Lorentzian base $(M_4,\hat g)$ and fiber $({\cal Y}, \bar g)$ is null geodesically complete if and only if $(M_4,\hat g)$ is null geodesically complete by itself. 
\end{theorem}

\proof 
According to \cite[Theorem 3.19(2)]{RS}, if $f \geq \epsilon >0$ it is enough that $\hat g/f^2$ be null geodesically complete to render the entire warped product (\ref{warped}) null geodesically complete too. If in addition $f\leq A$, then application of \cite[Theorem 2.2(i)]{CS} with
$\Omega= 1/f^2$ implies $\hat g/f^2$ is null geodesically complete if $\hat g$ is null geodesically complete.
\square

\section{Analysis of the conditions in the theorems}\label{sec:analysis}
Let us analyze the meaning and applicability of the main conditions in Theorem \ref{new1} and Theorem \ref{new2}.
First of all, observe that for any $\zeta_p\subset {\cal Y}_p$, there are always $\zeta_p$-orthogonal null geodesics with $n^i=0$ and thus with $\bar{N}^i(u)=0={\cal N}^i$ (those with $C=0$). These are null geodesics belonging to the 4-dimensional Lorentzian manifold $(M_4,\hat g)$. For these geodesics, the condition in Theorem \ref{new1} simplifies to (\ref{new1cond'}) as explained in Remark \ref{C=0}, while the negation of the condition in Theorem \ref{new2} (so that the spacetime cannot be null geodesically complete) reduces to 
\be
(C=0) \hspace{2mm} \Longrightarrow \hspace{3mm} \int_{\hat\gamma} \frac{1}{f} \frac{d^2 f}{du^2} du + \frac{1}{f(p)} n^a\partial_a f|_p <0 .\label{Ciszero}
\ee
It is obvious that (\ref{Ciszero}) supersedes (\ref{new1cond'}), as the latter obviously implies the former if $\zeta_p$ is future trapped (which means $\frac{1}{f(p)} n^a\partial_a f|_p <0$ for any $n^a$). Thus, we can concentrate on (\ref{Ciszero}), which becomes a necessary condition for the theorems to apply leading to null geodesic incompleteness. In more geometrical terms, taking into account that (with $C=0$) $\hat N^b\hat\nabla_b \hat N^a=0$, (\ref{Ciszero}) reads
\be
(C=0) \hspace{2mm} \Longrightarrow \hspace{3mm}  -\int_{\hat\gamma} \frac{1}{f} \hat{N}^a\hat{N}^b\hat\nabla_a\hat\nabla_b f > \left. \frac{1}{f} \hat{N}^a\hat\nabla_a f\right|_p \, .\label{Ciszero'}
\ee
Hence, if some extra dimensions start, say, contracting ---otherwise, one could similarly use the past version of the theorems--- along $M_4$-null directions (i.e. $\hat{N}^a\hat\nabla_a f |_p=n^a\partial_a f |_p < 0$) then it is enough that the Hessian of $f$ be non-positive on the corresponding null geodesics {\em on average}.

If condition (\ref{Ciszero'}) actually holds for \underline{all} null geodesics starting at a given ${\cal Y}_p$ (i.e. for a choice of $x^a =\hat x^a_p$) and if ${\cal Y}$ is compact, then this very ${\cal Y}_p$ acts as the submanifold leading to null geodesic incompleteness via the previous theorems. 

Observe that, from the expression of the Ricci tensor (\ref{ricci1}-\ref{ricci3}) and as $N^\mu =(\hat{N}^a,0)$ for these null geodesics with $C=0$, one has
$$
(C=0) \hspace{2mm} \Longrightarrow \hspace{3mm}  \frac{1}{n}\left(R_{\mu\nu} N^\mu N^\nu - \hat{R}_{ab} \hat{N}^a\hat{N}^b\right) = - \frac{1}{f}  \hat{N}^a\hat{N}^b\hat\nabla_a \hat\nabla_b f
$$
which can be further simplified in this situation to
$$
(C=0) \hspace{2mm} \Longrightarrow \hspace{3mm}  \frac{1}{n}\left(R_{ab}  - \hat{R}_{ab}\right)  \hat{N}^a\hat{N}^b= - \frac{1}{f}  \hat{N}^a\hat{N}^b\hat\nabla_a \hat\nabla_b f
$$ 
so that (\ref{Ciszero'}) can be rewritten in terms of the Ricci tensors 
\be
(C=0) \hspace{2mm} \Longrightarrow \hspace{3mm}   \frac{1}{n} \int_{\hat\gamma}\left(R_{ab}  - \hat{R}_{ab}\right)  \hat{N}^a\hat{N}^b > \left. \frac{1}{f} \hat{N}^a\hat\nabla_a f\right|_p \, .\label{Ciszero''}
\ee
This permits to analyze in greater detail when the necessary condition will hold (easy for instance if $\hat R_{ab} =\Lambda \hat{g}_{ab}$ on $(M_4,\hat{g})$). It is noticeable that the analysis can be performed in terms of the {\em null convergence condition} for the visible large 4-dimensional spacetime ($\hat R_{ab} \hat N^a\hat N^b\geq 0$) in comparison to that of the full spacetime {\em on null directions of the large 4-dimensional part} ($R_{\mu\nu}N^\mu N^\nu = R_{ab} N^a N^b \geq 0$ when  $C=0$). Another remarkable fact is that, even if some of the extra dimensions stay stationary, or expand while the others contract, there may be many situations where (\ref{Ciszero''}) also holds.

Consider, then, the case that the above necessary condition (\ref{Ciszero''}), or equivalently (\ref{Ciszero}), does not hold for {\em all} possible null geodesics orthogonal to {\em any} choice of ${\cal Y}_p$. Still, the condition can be satisfied by an appropriate subset of null geodesics with $C=0$, so that null geodesic incompleteness can still be derived from the theorems if those null geodesics are orthogonal to particular closed submanifolds $\zeta_p\subset {\cal Y}_p$. In this case, one still needs to check that the found inequality condition in the stronger Theorem \ref{new2} holds for the remaining null geodesics orthogonal to $\zeta_p$, those with $C>0$, and thus with $\bar{N}^i(u)\neq 0$.

In order to use again the negation of the condition in Theorem \ref{new2}:
\be
\int_{\gamma} \frac{1}{f^4}\delta^{AB}\overline{R}_{ijkl}{\cal N}^i {\cal N}^k \bar{E}^j_{A\parallel} \bar{E}^l_{B\parallel} -\bar{\theta}_{\bar{n}}
 >   (D-m) \left(\int_{\hat\gamma} \frac{1}{f} \hat N^b \hat\nabla_b(\hat N^a\hat\nabla_a f) + \frac{1}{f(p)} n^a\partial_a f|_p \right), \label{notnew2}
\ee
an analysis of the behaviour of $\hat N^b \hat\nabla_b(\hat N^a\hat\nabla_a f)$ along these null geodesics with $C>0$ is needed. The general expression for this second derivative is, on using (\ref{subgeod2}) 
\be
\hat N^b \hat\nabla_b(\hat N^a\hat\nabla_a f)= (C/f^3) \hat\nabla^b f \hat\nabla_b f +\hat{N}^a\hat{N}^b \hat\nabla_a\hat\nabla_b f . \label{DDf}
\ee
The first summand on the righthand side favors the singularity if the gradient of $f$ is non-spacelike: this is the case if the perturbation is truly dynamical, that is, if the dynamical part dominates over other possible accompanying perturbations. This should be assumed in what follows because, actually, in string theory keeping the values of the coupling constants and the Planck mass independent of position in space (to within experimental limits) requires that $f$ should essentially depend only on a time coordinate \cite{IU}, or in other words, that $\hat\nabla f$ is timelike and thus $\hat\nabla^b f \hat\nabla_b f <0$. Of course, there are also constraints in the time variation of such constants, but this is more easily accommodated within observational limits \cite{M,GV,KPW}, see \cite{U} for a review.

The last summand in (\ref{DDf}) can be analyzed as before, but taking into account that $\hat{N}^a$ are now timelike due to (\ref{C1}) with $C>0$. For instance, it is sufficient that the Hessian of $f$ be non-positive on the timelike curves $\hat\gamma(u)$ of $(M_4,\hat g)$ {\em on average} ---in fact, it would be enough to assume just that such Hessian is not too much positive on those timelike directions. A simple way to achieve this is to consider that $-\hat\nabla_a \hat\nabla_b f$, seen as a 2-index tensor, satisfies the {\em averaged} timelike convergence condition ---but again, much less is necessary. If such  a condition is assumed, then the necessary condition studied before for the $C=0$ geodesics will automatically follow by continuity. 

An important remark is that one only needs that the \underline{combination} of the two summands in  (\ref{DDf}) is non-negative, or not too much positive in comparison with the lefthand side in (\ref{notnew2}), {\em on average}.

Consider then the lefthand side in (\ref{notnew2}). As explained above, apart from the positive factor $f^{-4}$ this expression is an integral relative to the extra-dimensional space $({\cal Y},\bar{g})$ \underline{exclusively} ---and of course, $\bar{\theta}_{\bar n} $ is a quantity relative to the extra-dimensional space too. We remark that it is enough to find just \underline{one} compact submanifold $\zeta_p$ with the required property, and that the submanifold can have any dimension. Therefore, there are plenty of possibilities to play with, leading to a wider application of the singularity theorems. A few outstanding examples are
\begin{itemize}
\item
If the co-dimension is $5$, that is dimension $n-1$ so that $\zeta_p$ has co-dimension one as a submanifold of ${\cal Y}_p$, then $\delta^{AB}\bar{E}^j_{A\parallel} \bar{E}^i_{B\parallel}= \bar g^{ij}-{\cal N}^i{\cal N}^j/\bar g({\cal N},{\cal N})$ hence 
$$\delta^{AB}\bar{R}_{ijkl}{\cal N}^i{\cal N}^k \bar{E}^j_{A\parallel} \bar{E}^l_{B\parallel} =\bar{R}_{ij}{\cal N}^i{\cal N}^j
$$ 
and the condition (\ref{DDf}) simplifies accordingly involving the integral of $f^{-4}\bar{R}_{ij}{\cal N}^i{\cal N}^j $. In particular, the cases with Ricci-flat extra-dimensional space will be geodesically incomplete whenever the righthand side in (\ref{DDf}) is negative, as argued above, and if for instance $\bar\theta_{\bar n} \leq 0$. Compact minimal hypersurfaces in Ricci flat $({\cal Y},\bar g)$ are thus not compatible with geodesic completeness when dynamical warping functions (perturbations) have a negative righthand side in (\ref{DDf}). 
 \item
The other extreme case, if dim $\zeta_p =1$, i.e. a circle, are such that there is only one tangent vector and $$\delta^{AB}\bar{R}_{ijkl}{\cal N}^i{\cal N}^k \bar{E}^j_{A\parallel}\bar{E}^l_{B\parallel} =\bar{R}_{ijkl}{\cal N}^i {\cal N}^k \bar{E}^j_\parallel\bar{E}^l_\parallel
 $$
 is just the sectional curvature $K({\cal N},\bar E_\parallel)$ within $({\cal Y}_p,\bar g)$ along the projected $\bar\gamma$. Therefore, the existence of non-negative sectional curvatures {\em on average} can lead again to the inequality (\ref{DDf}) and therefore, to null geodesic incompleteness, under the same condition as in the previous case.
\item
For other co-dimensions, the integral on the lefthand side of (\ref{DDf}) is a sum of sectional curvatures along $\bar\gamma$, and therefore similar comments apply: existence of non-negative sectional curvatures {\em on average} is again the requirement.
\end{itemize}
In consequence, for dynamical perturbations with a timelike $\hat\nabla_a f$ the righthand side of (\ref{DDf}) will be negative in a large class of reasonable situations, and then one can find many Riemannian manifolds that, when attached to the spacetime via a warped product of type (\ref{warped}), will lead to null geodesic incompleteness, rendering the total spacetime unstable against these dynamical perturbations. Of particular interest may be examples where some field equations are fixed (observe that, hitherto, no field equations have ever been used so that all our results apply to generic theories based on a Lorentzian manifold). As happened with the Einstein static universe \cite{L1,L2,E}, in which case one uses Einstein's field equations for pressure-free matter to derive its dynamical instability, using field equations in the above setting may force the warping function to necessarily satisfy the conditions leading to singularities.

\section{Concluding remarks}\label{sec:conclusions}
We have substantiated, at least partially, the idea that extra spatial dimensions (compact or not) can be unstable in the sense that singularity theorems will apply and lead to null geodesic incompleteness. Allowing for {\em dynamical} perturbations ---functions $f$ with timelike gradient, for instance--- may not be safe. Hence, the analysis of the exact ``destroying power'' of such functions $f$ is of physical interest.

An important conclusion is that the generalized singularity theorems considerably broaden the situations where null geodesic incompleteness arises, providing support to arguments in \cite{CGHW} and indirectly to Penrose's \cite{P,P2}.


Here we have concentrated on the case where the dynamical perturbations cause the extra dimensions to start collapsing. The other case, when the extra dimensions start expanding, can be treated similarly, but to the past. It might then be argued that such a past geodesic incompleteness is not important, in the sense that we anyway expect something funny happening classically somewhere in the past ---the big bang. However, one should bear in mind that the expansion tends to endure, and thus the extra dimensions may eventually become very large, or even infinite, see also \cite{CGHW}.

On a positive side, the condition of the theorems, as given involving quantities of only the extra-dimensional space in comparison with well controllable quantities depending on $f$ and its derivatives, may help in finding the stable possibilities, providing information on which classes of extra-dimensional spaces $({\cal Y},\bar g)$ may be viable and why ---and for which warping functions $f$.

%

\section*{Acknowledgments}
NC was supported by the Belgian Interuniversity Attraction Pole P07/18 (Dygest) and by the KU Leuven Research Fund project 3E160361 ``Lagrangian and calibrated submanifolds''. JMMS is supported under Grants No. FIS2017-85076-P (Spanish MINECO/ AEI/ FEDER, UE) and No. IT956-16 (Basque Government). JMMS thanks G. Galloway for the initial discussion on this subject, which he raised in their conversations, and M. S\'anchez for many interesting comments and some clarifications concerning \cite{CS,RS}. Information provided by J.J. Blanco-Pillado has been very useful. JMMS is also grateful to Z. Wyatt for providing some references and to T. Harada and the Relativity group at Rikkyo University (Japan) for hospitality while this work was being completed.\\

\noindent\textbf{Author contributions statement:}
NC contributed to an earlier version of the present research and performed calculations needed for the material presented in the Appendix as part of her PhD thesis \cite{Cipriani}. JMMS conceived and proposed this research, developed the necessary methodology, is accountable for all the steps and calculations of the entire research process, and for structuring and writing the manuscript.

\section*{Appendix. Warped products: parallel transport}
\appendix

In this appendix we provide the general formulas for the parallel transport of vector fields in general warped product semi-Riemannian manifolds $\hat M\times_f {\cal Y}$
\be
g_{\mu\nu}dx^\mu dx^\nu= \hat{g}_{ab}(x^c)dx^a dx^b + f^2(x^c) \bar{g}_{ij}(x^k)dx^i dx^j  \label{warped1},
\ee
for the general case with arbitrary signatures and dimensions of the fiber and the base. As far as we know these results were first presented in \cite{Cipriani}.

Let $\gamma: x^\mu =x^\mu (u)$ be any parametrized curve (not necessarily a geodesic) with tangent vector $N^\mu :=dx^\mu (u)/du =(\hat{N}^a,\bar{N}^i)$. By definition, the parallel transport $\vec E(u)$ of any vector $\vec e$ along $\gamma$ is the vector field satisfying $N^\mu\nabla_\mu E^\nu =0$ along $\gamma$ which coincides with $\vec e$ initially, and therefore it is given by the unique solution to the system of ODEs
\be
\frac{d E^\mu}{du}+\Gamma^\mu_{\rho\sigma} N^\rho E^\sigma =0, \hspace{1cm}  E^\mu (0) =e^\mu. \label{PT}
\ee
In the coordinate system (\ref{warped1}) the connection symbols can be easily found to be
\bea
\Gamma^a_{ib} =0, \hspace{1cm}  \Gamma^i_{ab}=0, \hspace{1cm}  \Gamma^a_{bc} =\hat\Gamma^a_{bc},
\hspace{1cm} \Gamma^i_{jk}= \overline{\Gamma}^i_{jk}, \label{gamma1}\\
\Gamma^a_{ij} =- f \hat g^{ab}\partial_b f \bar g_{ij}, \hspace{1cm}  \Gamma^i_{aj} = \delta^i_j \frac{\partial_a f}{f},\hspace{1cm} \label{gamma2}
\eea
where $\hat\Gamma^a_{bc}$ and $\overline\Gamma^i_{jk}$ denote the connection symbols corresponding to $(\hat M,\hat g)$ and $({\cal Y},\bar g)$ respectively. Hence, (\ref{PT}) splits into
\bean
\frac{dE^a}{du} +\hat\Gamma^a_{bc} \hat N^b E^c - f \hat g^{ad}\partial_d f \bar g_{ij}\bar N^i E^j =0, \\
\frac{dE^i}{du} +\bar\Gamma^i_{jk}\bar N^j E^k +\delta^i_j  \frac{1}{f} \left(\bar N^j E^a\partial_a f +E^j \hat N^a\partial_a f\right)=0,
\eean
which can be appropriately rewritten as
\bea
\hat N^a\hat\nabla_a E^b &=& \bar{g}(\bar N,\bar E) f (\mbox{grad} f)^b , \label{DE1}\\
\bar N^i\overline\nabla_i E^j &=& -\frac{1}{f} \left(E^j  \hat N^a\partial_a f +\bar N^j E^a\partial_a f \right).\label{DE2}
\eea
Here and in what follows the notation $\hat E$ and $\bar E$ will refer to the respective projections of $\vec E(u)$. 
\subsection*{The case when $\gamma$ is a geodesic}
Assume now that $\gamma$ is an affinely parametrized geodesic, so that $N^\rho\nabla_\rho N^\mu =0$. Then, the scalar product $g(N,E)$ remains constant along $\gamma$ for any vector field $\vec E(u)$ which is parallel transported along $\gamma$. Thus, we can write
$$
g(\vec N,\vec E) = g(N(0),e) := a,
$$
and using (\ref{warped1})
$$
g(\vec N,\vec E) =\hat g(\hat N,\hat E) +f^2 \bar g(\bar N,\bar E)=a
$$
so that equation (\ref{DE1}) becomes in this situation
\be
\hat N^a\hat\nabla_a E^b = \left(a- \hat g(\hat N,\hat E) \right)\frac{1}{f} (\mbox{grad} f)^b \label{DE1geod}.
\ee
The importance of this equation is that this is a good transport equation relative to the base manifold $(\hat M,\hat g)$ exclusively. Thus, in order to find the parallel transported vector field $\vec E$ one can start by solving (\ref{DE1geod}) with initial condition $E^b (0)=e^b$ and, once its solution $E^b(u)$ is known, this can be inserted into (\ref{DE2}) which can then be solved to provide the remaining projection $E^i(u)$. 

Of particular interest to this paper is the case when furthermore the tangent vector $\vec{e}$ is initially orthogonal to $n^\mu := N^\mu(0)$. Then $\vec E(u)$ will be orthogonal to $\vec N$ all along $\gamma$ so that $a=0$. Hence, (\ref{DE1geod}) reduces now to
\be
\hat{N}^b\hat\nabla_b \hat{E}^a = - (\hat{g}_{bc}\hat{N}^b\hat{E}^c )\, \hat\nabla^a (\ln f) \label{DE1perp}
\ee
which, as before, is a well-defined transport equation along the projected curve $\hat\gamma$ within $(\hat M,\hat g)$. Notice, in particular, that if the initial condition entails $E^a(0) = e^a =0$ then the unique solution is given by $E^a(u)=0$ for all $u$. 

Once the solution of (\ref{DE1perp}) is explicitly known, one can solve the remaining system (\ref{DE2}). In the present case, this can be done fully explicitly and the solution is given by 
\be
\bar{E}^i = fh \bar{N}^i +\frac{1}{f} \bar{E}^i_{\parallel},\label{barE}
\ee
where $\bar{E}^i_{\parallel}$ is the parallel transport of $f(0) e^i$ in $({\cal Y},\bar g)$ along the projected curve $\overline{\gamma}$ defined by $ x^i(u)$:
$$
\bar{N}^j\overline\nabla_j \bar{E}^i_{\parallel} =0, \hspace{1cm} \bar{E}^i_\parallel (0) =f(0) e^i,
$$
while $h(u)$ is the unique solution of 
\be 
\frac{dh}{du} = \hat{E}^a\partial_a (1/f), \hspace{1cm} h(0)=0.\label{dh}
\ee

To end this appendix, let us consider two special cases of prominent interest, depending on whether the initial vector $\vec e$ has vanishing projection to either the base or the fiber ---while keeping the orthogonality condition $g(\vec n,\vec e)=0$.
\begin{itemize}
\item Case with $e^\mu =(0,e^i)$, so that $e^a=0$. In this case, as explained above, the unique solution to (\ref{DE1perp}) is $E^a(u)=0$. Hence, equation (\ref{dh}) becomes trivial with unique solution $h(u)=0$ and thus the final solution for the parallel transport vector field is
$$
E^\mu (u) = (0, \bar E^i_\parallel /f) .
$$
This is the solution used in the main text.
\item Case with $e^\mu =(e^a,0)$, so that $e^i =0$. In this case $\bar E^i_\parallel =0$ along $\gamma$ and (\ref{barE}) simplifies to
$$
\bar{E}^i = fh \bar{N}^i
$$
with $h$ the solution of (\ref{dh}). Using this, the orthogonality condition along $\gamma$ becomes
$$
0=g(\vec N,\vec E)=\hat g(\hat N,\hat E) +f^2 \bar g(\bar N,fh\bar N)=\hat g(\hat N,\hat E) +f^3h \bar g(\bar N,\bar N),
$$
which can be introduced into (\ref{DE1perp}) to get
$$
\hat{N}^b\hat\nabla_b \hat{E}^a = h \bar g(\bar N,\bar N) f^2 \hat\nabla^a f .
$$
But along any geodesic (not necessarily null) formula (\ref{C}) holds \cite{O} where now $C$ can have any sign, 
so that the previous equation reads
$$
\hat{N}^b\hat\nabla_b \hat{E}^a =C\frac{h}{f^2}\hat\nabla^a f  =- C h \hat\nabla^a(1/f).
$$
From these calculations follows in particular that ($b$ is a constant)
$$
b:= g(\vec e,\vec e) =g(\vec E,\vec E)= \hat g(\hat E,\hat E)+f^2 \bar g(\bar E,\bar E) =\hat g(\hat E,\hat E)+Ch^2,
$$
as well as 
$$
\hat g(\hat N,\hat E)=-C \frac{h}{f}.
$$
In the case that $\gamma$ {\em is a null geodesic} so that $g(\vec N,\vec N)=0$ then (\ref{C1}) also holds 
and all the above suggests to define 
\be
\hat{{\cal E}}^a := \hat{E}^a -f h \hat{N}^a, \label{calE}
\ee
which is a vector field along the {\em null} $\gamma$ satisfying
$$
\hat g(\hat {\cal E}, \hat {\cal E}) =b, \hspace{5mm} \hat g(\hat N,\hat {\cal E})=0, \hspace{5mm} \hat N^b\hat\nabla_b \hat {\cal E}^a =-\frac{d}{du} (hf) \hat N^a, \hspace{5mm} \hat {\cal E}^a(0)=e^a .
$$
From this, a straightforward calculation allows one to show that, whenever $C\neq 0$ so that the projected curve $\hat\gamma$ is non-null in $(\hat M,\hat  g)$, the vector field $\hat{\cal E}$ is {\em Fermi-Walker} transported \cite{MTW} along the projected curve $\hat{\gamma}$. We believe this is a new result.
\end{itemize}

\end{document}